\documentclass[pra
	       ,nofootinbib
	       ,floatfix
	       ,superscriptaddress
	       ,twocolumn
	       ]{revtex4-1}

\usepackage{graphicx}
\usepackage{dcolumn}
\usepackage{bm}
\usepackage[utf8]{inputenc}
\usepackage{amssymb}
\usepackage{amsmath}
\usepackage{pstricks}
\usepackage{color}
\usepackage{slashed}
\usepackage{braket}
\usepackage{hyperref}
\usepackage{natbib}
\usepackage{extarrows}
\usepackage{titlesec}
\usepackage{wasysym}
\usepackage[normalem]{ulem}

\definecolor{ao(english)}{rgb}{0.0, 0.5, 0.0}
\definecolor{applegreen}{rgb}{0.55, 0.71, 0.0}
\definecolor{cadetblue}{rgb}{0.37, 0.62, 0.63}
\definecolor{cadet}{rgb}{0.33, 0.41, 0.47}
\definecolor{byzantine}{rgb}{0.74, 0.2, 0.64}

\newcommand{\eq}[1]{(\ref{eq:#1})}
\newcommand{\Eq}[1]{Eq.~(\ref{eq:#1})}

\newcommand{\Fig}[1]{Fig.~\ref{fig:#1}}

\newcommand{\FigED}[1]{appendix Fig.~\ref{fig:#1}}
\newcommand{\FigsED}[2]{appendix Figs.~\ref{fig:#1}--\ref{fig:#2}}

\newcommand{\nK}{\, \mathrm{nK}}
\newcommand{\Hz}{\, \mathrm{Hz}}
\newcommand{\kHz}{\, \mathrm{kHz}}
\newcommand{\kHzpms}{\, \mathrm{kHz}/\mathrm{ms}}
\newcommand{\ms}{\, \mathrm{ms}}

\newcommand{\imum}{\, \mu \mathrm{m}^{-1}}

\makeatletter
\newcommand*{\balancecolsandclearpage}{%
  \close@column@grid
  \clearpage
  \twocolumngrid
}
\makeatother

\makeatletter
\let\cat@comma@active\@empty
\makeatother

\begin{document}

\title{Observation of universal dynamics in an isolated one-dimensional Bose gas far from equilibrium}

\author{Sebastian Erne}
\affiliation{Vienna Center for Quantum Science and Technology, Atominstitut, TU Wien, Stadionallee 2, 1020 Vienna, Austria}
\affiliation{Institut f\"ur Theoretische Physik,
             Ruprecht-Karls-Universit\"at Heidelberg,
             Philosophenweg~16,
             69120~Heidelberg, Germany}
\affiliation{Kirchhoff-Institut f\"ur Physik,
             Ruprecht-Karls-Universit\"at Heidelberg,
             Im Neuenheimer Feld 227,
             69120~Heidelberg, Germany}
\affiliation{School of Mathematical Sciences, University of Nottingham, University Park, Nottingham NG7 2RD, UK}
\affiliation{Centre for the Mathematics and Theoretical Physics of Quantum Non-Equilibrium Systems, \protect\\ University of Nottingham, Nottingham NG7 2RD, UK}

\author{Robert B\"ucker}
\affiliation{Vienna Center for Quantum Science and Technology, Atominstitut, TU Wien, Stadionallee 2, 1020 Vienna, Austria}
\affiliation{Max Planck Institute for the Structure and Dynamics of Matter, Luruper Chaussee 149, 22761 Hamburg, Germany}

\author{Thomas Gasenzer}
\affiliation{Institut f\"ur Theoretische Physik,
             Ruprecht-Karls-Universit\"at Heidelberg,
             Philosophenweg~16,
             69120~Heidelberg, Germany}
\affiliation{Kirchhoff-Institut f\"ur Physik,
             Ruprecht-Karls-Universit\"at Heidelberg,
             Im Neuenheimer Feld 227,
             69120~Heidelberg, Germany}
             
\author{J\"urgen Berges}
\affiliation{Institut f\"ur Theoretische Physik,
             Ruprecht-Karls-Universit\"at Heidelberg,
             Philosophenweg~16,
             69120~Heidelberg, Germany}

\author{J\"org Schmiedmayer} 
\email{schmiedmayer@atomchip.org}
\affiliation{Vienna Center for Quantum Science and Technology, Atominstitut, TU Wien, Stadionallee 2, 1020 Vienna, Austria}

\begin{abstract}
 We provide experimental evidence of universal dynamics far from equilibrium during the relaxation of an isolated one-dimensional Bose gas. Following a rapid cooling quench, the system exhibits universal scaling in time and space, associated with the approach of a non-thermal fixed point. The time evolution within the scaling period is described by a single universal function and scaling exponent, independent of the specifics of the initial state. Our results provide a quantum simulation in a regime, where to date no theoretical predictions are available. This constitutes a crucial step in the verification of universality far from equilibrium. If successful, this may lead to a comprehensive classification of systems based on their universal properties far from equilibrium, relevant for a large variety of systems at different scales.
\end{abstract}

\date{\today}

\maketitle

Understanding isolated quantum systems far from equilibrium and the question about the subsequent thermalisation process concerns one of the most pressing open problems in quantum many-body physics \cite{Polkovnikov2011a,Gogolin2016}. There is strong theoretical evidence that sufficiently far from equilibrium a wide variety of systems exhibit universal scaling during their evolution,  independent of the details of their initial state and microscopic properties \cite{Berges:2008wm,Schole:2012kt,Berges:2014bba,svistunov1991highly,Micha2003a,BAIER200151,PhysRevD.89.074011,Schmidt:2012kw,Moore:2015adu,Orioli:2015dxa,Chantesana2018arXiv180109490C,Deng:2018xsk}. However, experimental evidence is still missing. Here, we report universal scaling in time and space following a strong cooling quench transferring a 3D ultra-cold Bose gas into a one-dimensional quasicondensate. In the scaling regime, the time evolution of the system is found to be described by a time independent universal function and a single scaling exponent in the infrared. The non-equilibrium evolution features the transport of an emergent conserved quantity in the scaling region, finally leading to the build-up of a quantum degenerate quasicondensate. Our results establish universal scaling dynamics in an isolated quantum many-body system, providing conceptually new access to time evolution far from equilibrium relevant for a large variety of systems \cite{Orioli:2015dxa,Berges:2014bba}.

%
Relaxation and thermalisation are characterised by loss of information about the details of the initial state. The unitary quantum evolution of isolated systems, however, preempts any such loss of information on a fundamental level. To resolve this contradiction, it is generally assumed that the complexity of the involved many-body states and their dynamics leads to an insensitivity to details about the initial state for any realistic observable \cite{Polkovnikov2011a,Gogolin2016}. Consequently, at late times, the system can be characterised by few conserved quantities only. 

An enhanced loss of details about the underlying microscopic physics is known to occur, e.g., for critical scaling near phase transitions \cite{Hohenberg1977a,Janssen1992a}, for the phenomenon of ageing \cite{Calabrese2005a.JPA38.05.R133} or for coarsening \cite{Bray1994a.AdvPhys.43.357}. Characterising these (close to equilibrium) systems through their universal scaling properties has lead to tremendous advances in the understanding of complex quantum many-body systems \cite{Hohenberg1977a,ZinnJustin2004a}. 
Little is known if and how the effective information loss during the time evolution starting from a general far-from-equilibrium state can be connected to universality, even away from any phase transition.

\begin{figure*}[t!]
\includegraphics[width=1\textwidth]{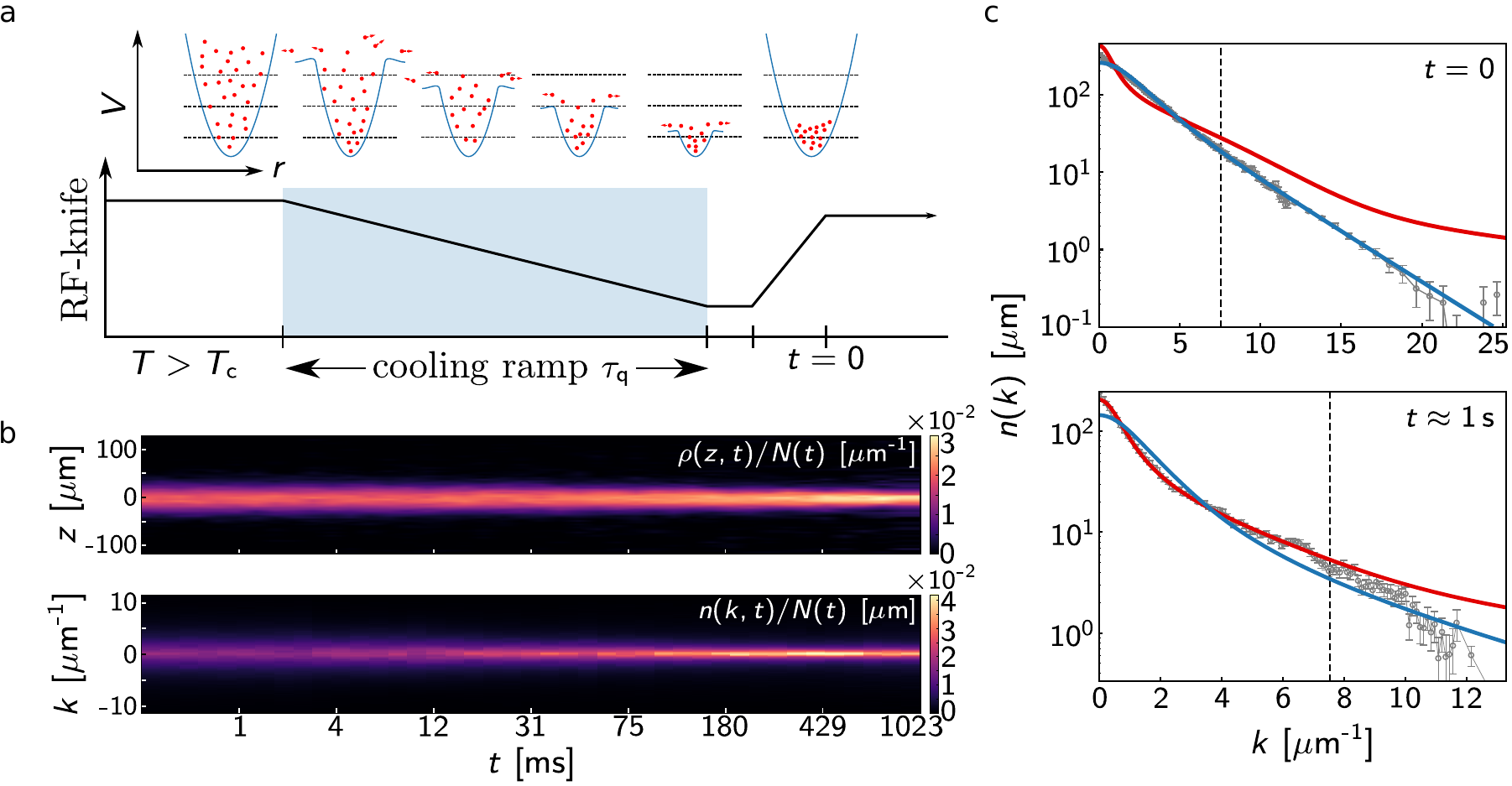}
\caption{ {\bf Experimental scheme and overview of results.} 
{\bf a,} schematics of the experimental cooling quench. During the quench the trap depth is linearly ramped to its final value within a time $\tau_\mathrm{q} \approx 7 \, \mathrm{ms}$. The final value lies below the first radially excited state (indicated by the dashed lines). The trap depth is subsequently held at its final position for $\approx 0.5 \, \mathrm{ms}$ and then within $\approx 1 \, \mathrm{ms}$ raised  again leading to a far-from-equilibrium one-dimensional Bose gas.
{\bf b,} time evolution of the density $\rho(z,t)$ (upper panel) and of the single-particle momentum distribution $n(k,t)$ (lower panel). Each distribution is normalised to the time-dependent atom number $N(t)$.
{\bf c,} initial (upper panel) and final (lower panel) momentum distributions. The data for high momenta is binned over $7$ adjacent $k$ values to lower the noise level. Error bars are marking the standard error of the mean. The solid blue and red lines are theoretical fits of the random-soliton model and of a thermal quasicondensate, respectively (see \FigED{ED_fitresults}). The vertical dashed line corresponds to the momentum of the first radially excited state.
}
\label{fig:fig1}
\end{figure*}

It has recently been proposed that isolated systems far from equilibrium exhibit scaling in time and space associated to non-thermal fixed points \cite{Berges:2008wm,Schole:2012kt,Berges:2014bba,Orioli:2015dxa}. There is growing theoretical evidence for non-thermal universality classes encompassing both relativistic as well as non-relativistic systems~\cite{Berges:2014bba,Orioli:2015dxa}. These non-equilibrium attractor solutions require, in contrast to equilibrium critical phenomena, no fine-tuning of parameters. Unlike the phenomenon of prethermalisation to quasi-stationary states~\cite{Berges:2004ce,Gring2012}, which are approximately time-translation invariant, the non-thermal scaling solutions describe a time-dependent evolution.

In the present work, we experimentally study the dynamics following a strong cooling quench and identify a time window during which the system shows universal behavior far from equilibrium. 

\begin{figure*}[t!]
\includegraphics[width=1\textwidth]{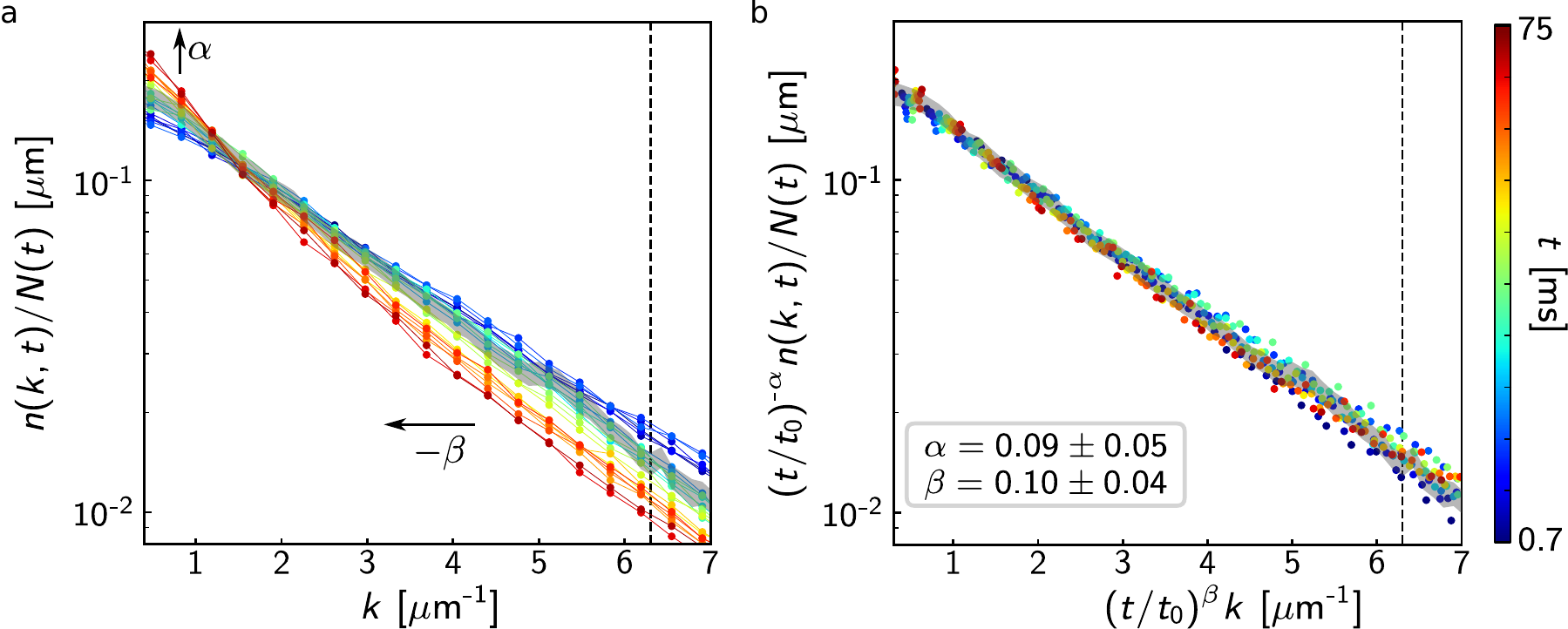}
\caption{{\bf Universal scaling dynamics.}
{\bf a,} Time evolution of the measured momentum distributions. For better visibility, the data is binned over $3$ adjacent points in momentum space, with the time encoded in colors (from blue to red). The gray line indicates the reference distribution at $t_0=4.7 \, \mathrm{ms}$, its width depicts the $95\%$ confidence interval at $t_0$. The vertical dashed line limits the scaling region in the UV.
{\bf b,} Momentum distribution rescaled according to \Eq{ScalingFunction}. Depicted as a rescaled function $(t/t_0)^{-\alpha} n(\tilde{k},t)$ of the rescaled variables $\tilde{k}=(t/t_0)^\beta k$ the data for all times collapse to a single curve, representing the distribution at the reference time $t_0$. The exponents $\alpha = 0.09 \pm 0.05$ and $\beta = 0.10 \pm 0.04$ are determined via the maximum likelihood function.
}
\label{fig:fig2}
\end{figure*}

We start our experiment with a thermal gas of ultra-cold $^{87}$Rb in an extremely elongated quasi one-dimensional (1D) harmonic trap (transverse confinement $\omega_\perp = 2 \times 10^4 \, \mathrm{s}^{-1}$, longitudinal confinement $\omega_\parallel = 30 \, \mathrm{s}^{-1}$) just above the critical temperature. In the final cooling step, the trap depth is lowered fast compared to the longitudinal thermalisation time scale (see \Fig{fig1}a). This leads to a rapid removal of high-energy atoms, predominantly in the radially excited states, and hence constitutes an almost instantaneous cooling quench of the system. At the end of the cooling ramp the trap depth lies below the first radially excited energy level and only longitudinal excitations remain. After a short holding period of 1 ms, allowing the atoms with large transverse energy to leave, we rapidly increase the trap depth. In this way we prepare an isolated far-from-equilibrium 1D system. The gas is then left to evolve in this deep potential for variable times $t$ up to $\approx 1 \, \mathrm{s}$, during which the universal scaling dynamics takes place. 

We probe the system's evolution through two sets of measurements (see Methods for details): The in-situ density distribution $\rho(z,t)$ is measured by standard absorption imaging \cite{AduSmith2011a} after a short time of flight of $t_\mathrm{tof} = 1.5\,\mathrm{ms}$, whereby the expansion is predominantly along the tightly confined radial directions. The momentum distribution $n(k,t)$ of the trapped gas is measured after a long time of flight of $t_\mathrm{tof} = 46\,\mathrm{ms}$  through single-atom resolved fluorescent imaging in a thin light sheet \cite{Buecker2009a}. For each hold time $t$ the distributions are averaged over many independent measurements (see Methods).

A typical time evolution of these profiles is shown in \Fig{fig1}b. The far-from-equilibrium state at early times exhibits strongly broadened density and momentum distributions. In the beginning the momentum distribution $n(k)$ follows a characteristic exponential decay $n(k) \sim \exp(-k \xi_\mathrm{s})$ for large values of $k$. At late times the system is well described by a thermal quasicondensate (see \Fig{fig1}c and Methods and SI for details), revealing its relaxation to thermal equilibrium (see \FigED{ED_fitresults}). The momentum distribution is then only determined by the Lorentzian function, its width given by the thermal coherence length $\lambda_T = 2 \hbar^2 \rho(z) / m k_\mathrm{B} T$. During the evolution a clear peak emerges at low momenta, signaling the quasicondensation of the system in momentum space. In the following we analyse the thermalisation process, providing the link between the far-from-equilibrium state at early times and the final equilibrium state observed.  

For the initial state of the far-from-equilibrium evolution we find $n(k)$ in good agreement  with a theoretical model of randomly distributed solitonic defects (RDM) \cite{Schmidt:2012kw} (see \Fig{fig1}c). At low momenta the RDM has a Lorentzian shape $n(k) \sim [1+(k/n_\mathrm{s})^2]^{-1}$, its width defined by the defect density $n_\mathrm{s}$.  At high momenta $n(k)$ exhibits a characteristic exponential decay $n(k) \sim \exp(-k \xi_\mathrm{s})$, determined by the width $\xi_\mathrm{s}$ of the localised density suppression associated with a solitonic defect. 

Since we probe the system immediately after the almost instantaneous quench, these defects are not equilibrated (see \FigED{ED_fitresults}), showing a reduced defect width $\xi_\mathrm{s} = 0.07 \, \mu\mathrm{m} \approx \xi_\mathrm{h} / 3$, with the peak healing length $\xi_\mathrm{h} = \hbar / \sqrt{2 m g_\mathrm{1D} n_0}$ determining the equilibrium width of a soliton, and a very high density $n_\mathrm{s} = 1.4 \, \mu \mathrm{m}^{-1}$. While the nucleation of solitons is predicted by the Kibble-Zurek mechanism \cite{del2014universality}, the almost instantaneous quench here creates an initial state with a strong overpopulation of high energy modes. This very far-from-equilibrium state sets the initial conditions for the subsequent thermalisation process and facilitates the observation of the emerging universal dynamics during the relaxation of the system.

\begin{figure}[t!]
\includegraphics[width=0.47\textwidth]{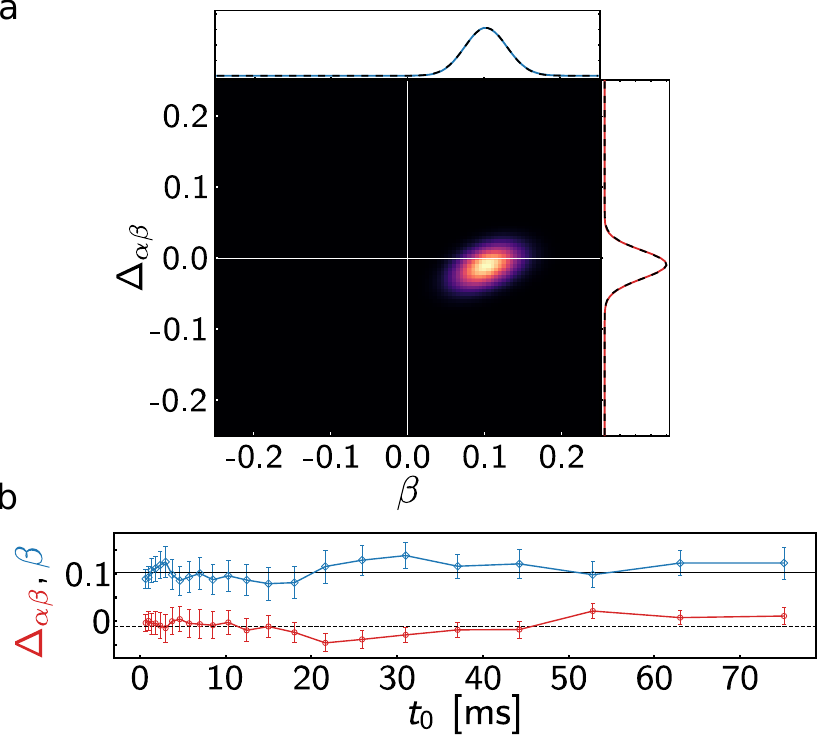}
\caption{{\bf Scaling exponents.}
{\bf a,} The combined two-dimensional likelihood function, averaged over all times $t$ and reference times $t_0$ within the scaling period, for three different initial conditions reveals a clear peak which yields the non-vanishing scaling exponents $\alpha \approx \beta = 0.1 \pm 0.03$ with a deviation between the two exponents of $\Delta_{\alpha \beta}=\alpha-\beta=-0.01 \pm 0.02$. The error is estimated using a Gaussian fit to the marginal likelihood functions (see sides). {\bf b,} Dependence of the scaling exponents on the reference time $t_0$. The exponents are, to a good approximation, independent of $t_0$ and agree well with the mean predictions. The errors denote the standard deviation obtained by a Gaussian fit to the marginal likelihood functions at each reference time separately.
}
\label{fig:fig3}
\end{figure}

The time evolution of the normalised momentum distribution $n(k,t)/N(t)$ is shown in \Fig{fig2}a for the first 75 ms following the quench. The distribution function shifts with time towards lower momentum scales while the occupancy is growing in the infrared. In general, $n(k,t)$ depends on $k$ and $t$ separately. 

However, it has been suggested \cite{Orioli:2015dxa} that overpopulated fields far from equilibrium can give rise to  universal behaviour signaled by the infrared scaling property of the distribution functions 
\begin{align}
   n(k,t) =  (t/t_{0})^{\alpha}\,f_\mathrm{S}([t/t_{0}]^{\beta}k) ~,
   \label{eq:ScalingFunction}
\end{align}
where $t_{0}$ denotes an arbitrary reference time within the period where $n(k,t)$ shows the scaling behavior.

\Fig{fig2}b shows that indeed one can find scaling exponents $\alpha$ and $\beta$ such that, in the infrared, the rescaled distributions $(t/t_0)^{-\alpha} n(\tilde{k},t)$ as a function of the rescaled momenta $\tilde{k} = (t/t_0)^\beta k$ collapse to a single curve $f_\mathrm{S}(\tilde{k})=n(\tilde{k},t_0)$. This indicates that below a characteristic momentum scale $k_\mathrm{S}$ the distribution function $n(k,t)$ depends on space and time only through scaling of a single universal function $f_\mathrm{S}(\tilde{k})$. The scaling exponents are found to be $\alpha = 0.09 \pm 0.05$ and $\beta = 0.1 \pm 0.04$, which indicates $\alpha \simeq \beta$ (see Methods for details on the error estimation).

We demonstrate the predicted insensitivity of the universal properties to details of the initial state by comparing the evolution for different initial conditions prior and posterior to the cooling quench. We find excellent agreement for the scaling exponents, obtained independently through a scaling analysis for each of the three measurements (see \FigsED{ED_rescaling}{ED_global_observables}). This shows the generality and robustness of these nonequilibrium attractor solutions, as, in contrast to equilibrium critical phenomena where the temperature has to be adjusted to observe scaling, no fine-tuning of parameters is required.

\begin{figure*}[t]
\includegraphics[width=1\textwidth]{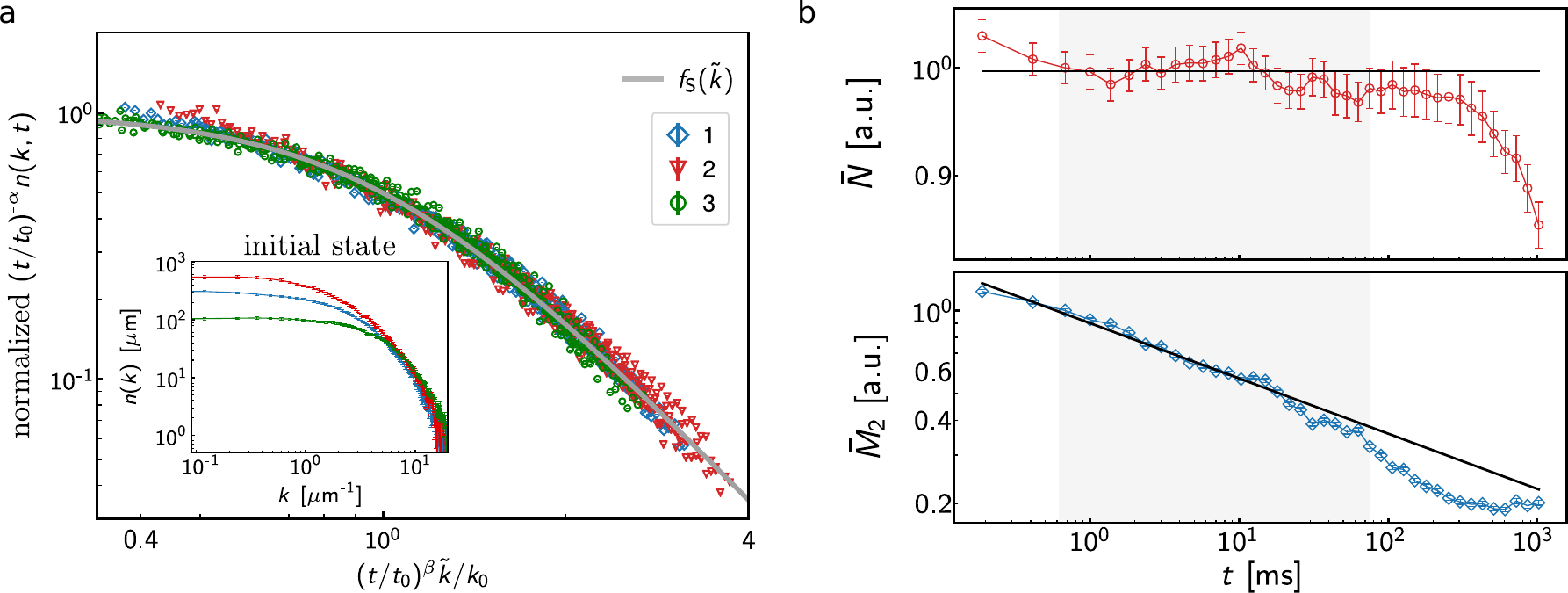}
\caption{{\bf Universal scaling function and spatially averaged observables.}
{\bf a,} Universal scaling function for varying initial conditions $N=1700 \, , n_\mathrm{s}=1.4 \, \mu \mathrm{m}^{-1}$ (blue), $N=2800 \, , n_\mathrm{s}=0.9 \, \mu \mathrm{m}^{-1}$ (red), and $N=1150 \, , n_\mathrm{s}=2.3 \, \mu \mathrm{m}^{-1}$ (green). All initial conditions collapse to a single universal function $f_\mathrm{S}$ with exponent $\zeta = 2.39 \pm 0.18$ (gray solid line) for all times within the scaling region. The rescaled experimental data is binned over $3$ adjacent points in $k$ for clarity. The small deviations for low momenta are due to the finite expansion time of the gas (see Methods). The initial single-particle momentum distribution at the end of the quench is depicted in the inset.
{\bf b,} Scaling of averaged observables. The fraction of particles in the scaling region $\bar{N} \sim (t/t_0)^{\Delta_{\alpha \beta}}$ (upper panel) becomes approximately conserved (solid black line) within the scaling period (gray shaded region) while being transported towards the IR. Scaling of the mean kinetic energy per particle in the scaling region, $\bar{M}_2 \sim (t/t_0)^{-2 \beta}$, (lower panel) clearly signals the extent of the scaling period in time through deviations from the predicted scaling (solid black line). The error bars are marking the $95\%$ confidence interval.
}
\label{fig:fig4}
\end{figure*}

The universal character allows us to directly relate the predictions for each measurement, resulting in the combined likelihood function presented in \Fig{fig3}a. We consider, for the analysis, the approximately uncorrelated exponents $\beta$ and $\Delta_{\alpha \beta} = \alpha - \beta$. In agreement with each individual measurement, we find the clearly non-vanishing exponent $\beta = 0.1 \pm 0.03$ and within errors a vanishing exponent $\Delta_{\alpha \beta} = -0.01 \pm 0.02$, and thus $\alpha = 0.09 \pm 0.03$. The expected independence of the scaling exponents $\alpha$, $\beta$ on the reference time $t_0$ is shown in \Fig{fig3}b.

We further demonstrate that the shape of the scaling function $f_\mathrm{S}(\tilde{k})$ is universal. As shown in \Fig{fig4}a the data for three different initial conditions follow a single universal function $f_\mathrm{S}(\tilde{k})$ for all times during which the system shows scaling dynamics. For the scaling function we consider the form \mbox{$f_\mathrm{S} \sim [1+(\tilde{k} / k_0)^{\zeta}]^{-1}$} \cite{Orioli:2015dxa,Karl2017b.NJP19.093014}, where the exponent $\zeta = 2.39 \pm 0.18$ is obtained through a single maximum likelihood fit to all experimental realisations simultaneously. For a fixed exponent the non-universal scales, i.e.\ the norm and momentum scale $k_0$ rescaling the dimensionless momentum $\tilde{k}/k_0$, are determined through a least-square fit for each experimental realisation (see Methods).
The shape of the momentum distribution within the scaling period exhibits marked differences to the thermal distribution (cf.~\Fig{fig1}c and \FigED{ED_fitresults}), clearly indicating a non-thermal scaling phenomenon.

The extent of the scaling region in time is visible from the scaling behavior of the spatially averaged observables $\bar{N}$ and $\bar{M}_2$ (see Methods). These describe the fraction of particles and the mean energy per particle in the time dependent scaling region $|k| \leq (t/t_0)^{- \beta} k_\mathrm{S}$ in momentum space, respectively. Based on the scaling Ansatz \Eq{ScalingFunction} we find $\bar{N} \sim (t/t_0)^{\Delta_{\alpha \beta}}$ and hence, as $\Delta_{\alpha \beta} \approx 0$, the emergence of a conserved quantity. This is confirmed in \Fig{fig4}b where $\bar{N}$ is approximately constant in the scaling period, while it shows a clear time dependence before and afterwards. 

The values for the scaling exponents $\alpha$ and $\beta$ determine the direction and speed with which the particles are being transported. Since these values are positive, a given momentum $k$ in this regime scales as $k/k_0\sim t^{-\beta}$ such that the transport is directed towards the infrared. This transport of particle number leads finally to the observed build-up of the quasicondensate and the approach to thermal equilibrium at late times. We further note that the mean energy shows a power-law behavior $\bar{M}_2 \sim (t/t_0)^{-2 \beta}$ and is in accordance with the determined scaling exponent $\beta$. Therefore, while the particle number in the scaling region is conserved, energy is transported outside this region to higher momenta. Based on the scaling properties of these global observables we can readily identify the scaling period to include the times $t \approx 0.7 \dots 75 \, \mathrm{ms}$.

The far-from-equilibrium universal scaling dynamics in isolated Bose gases following a strong cooling quench or for equivalent initial conditions has been studied theoretically by means of non-perturbative kinetic equations \cite{Orioli:2015dxa,Chantesana2018arXiv180109490C}. Therein, the universal scaling function is expected to depend on dimensionality $d$. The predicted power-law fall-off $n(k)\sim k^{-\zeta}$, with $\zeta=d+1$ \cite{Chantesana2018arXiv180109490C}, is consistent with the approximate form of the scaling function given by the RDM and the quasicondensate at low momenta but differs (slightly) from the experimental results. 

More prominently, a scaling analysis of the kinetic quasiparticle transport yields the exponent $\beta=1/2$ \cite{Orioli:2015dxa} in \Eq{ScalingFunction} to be independent of $d$. This theory, however, is not expected to fully apply. In particular, in $d=1$, due to the kinematic restrictions from energy and momentum conservation, the associated transport is expected to vanish.

The contributions of the higher dimensions to the 1D physics are a plausible path to explain the non standard scaling function and scaling exponents observed. Initially there is a small population of atoms with momenta large enough to excite thermalising collisions \cite{Mazets2008c}, and a very small initial seed can lead to thermalisation as observed in \cite{Li2018}. This is confirmed by a quasicondensate fit to the final momentum distribution where, assuming thermal equilibrium, one obtains an excited state population of 11\% ($T = 95 \, \mathrm{nK} = 0.6 \hbar \omega_\perp$). Our experimental results provide a quantum simulation near the dimensional crossover between 1D and 3D physics, establishing universal scaling dynamics far-from-equilibrium in a regime where currently no theoretical predictions are available.

The presented direct experimental evidence of scaling dynamics in an isolated far-from-equilibrium system presents a crucial step towards a description of non-equilibrium evolution by non-thermal fixed points and the associated phenomena. Similar exciting phenomena have recently been observed by the Oberthaler group \cite{Pruefer2018} in Heidelberg in a Spin-1 system, however with scaling exponent $\beta = 1/2$.
If successful, such programs can lead to a unified description of non-equilibrium evolution reminiscent of the classification of equilibrium critical phenomena in terms of renormalization group fixed points \cite{Wilson1975a.RevModPhys.47.773,Kadanoff1990a}. This may lead to a comprehensive classification of systems based on their universal properties far from equilibrium, relevant for a large variety of systems at different scales.

\vspace*{0.3cm}
\noindent{\bf Acknowledgements:} We thank J.~Brand, L.~Carr, M.~Karl, P.~Kevrekidis, P.~Kunkel, D.~Linnemann, A.N.~Mikheev, B.~Nowak,  M.K.~Oberthaler, J.M.~Pawlowski, A.~Pi\~{n}eiro Orioli, M.~Pr\"ufer,  W.~Rohringer, C.M.~Schmied,  M.~Schmidt, J.~Schole, H.~Strobel for discussions on the topics described here. We thank T.~Berrada, S.~van Frank, J.-F.~Schaff, T.~Schumm for help with the experiment during data taking. This work was supported by the SFB 1225 'ISOQUANT' and Grant No.~GA677/7,8 financed by the German Research Foundation (DFG) and Austrain Science Fund (FWF), the ERC advanced grant QuantumRelax, the Helmholtz Association (HA216/EMMI), the EU (FET-Proactive grant AQuS, Project No. 640800), and by Heidelberg University (CQD). S.E.~acknowledges partial support through the EPSRC Project Grant (EP/P00637X/1). J.S., J.B., and T.G.~acknowledge the hospitality of the Erwin Schr\"odinger Institut in the framework of their thematic program \emph{Quantum Paths}.

\bibliography{universal_1d_arXiv}


\clearpage
\appendix
\section*{Appendix}

\subsection{Preparation of the gas and cooling quench}

The initial thermal Bose gas is prepared using our standard procedure to produce ultra-cold gases of ${}^{87}$Rb on an atom chip~\cite{Reichel}. We prepare a thermal cloud of typically $N = (2.7 - 3.2) \cdot 10^4$ atoms initially in an elongated, $\omega_{\parallel} = 2\pi \cdot 23\Hz$ and $\omega_{\perp} = 2\pi \cdot 3.3\kHz$, deep trapping potential $V_\mathrm{i} \approx h \cdot (130 - 160) \kHz$ at a temperature $T \approx 530 - 600 \nK$. The atoms are held in this configuration for $100 \ms$ to ensure a well-defined initial state. The thermal cloud is both above the dimensional crossover to an effective one-dimensional system and the critical temperature $T_\mathrm{c}$ for the phase transition to a 3D Bose-Einstein condensate, and therefore has a large excess of particles in transversally excited, high-energy states. The trap depth is reduced to its final value $V_\mathrm{f}$ at a constant rate $R_\mathrm{q} = (V_\mathrm{i} - V_\mathrm{f}) / \tau_\mathrm{q} = h \cdot 25 \kHzpms$, by applying radio-frequency (RF) radiation at a time dependent frequency (RF-knife), leading to an energy-dependent transition of atoms from a trapped to an un-trapped spin state. This allows the high-energy particles to rapidly leave the trap, leading to the competing time scales $\tau_q$ and the typical collision times needed for re-equilibration of the system. The final trap depth is $V_\mathrm{f} \approx h \cdot 2 \kHz$, which lies below the first radially excited state of the trapping potential $h V_\mathrm{f} < \hbar \omega_{\perp}$. At the end of the cooling ramp, the RF-knife is held at its final position for $\tau_\mathrm{h} = 0.5 \ms$ before it is faded out within $\tau_\mathrm{f} \approx 1 \ms$, thereby raising the trap depth to $V \approx h \cdot 20 \kHz$. Additionally, since the RF-knife slightly reduces the radial trapping frequency, this leads to a small interaction quench ($\sim 10\%$) of the 1D system. The system is therefore rapidly quenched to the quasi one-dimensional regime, finally only occupying the transverse ground state. The experimental realisations 1 to 3 reported in the main text have final atom numbers $N \approx 1700 \, , \, 2800 \, , \, 1150$ and agree well with the RDM with a defect density $n_\mathrm{s} = 1.4 \, , \, 0.9 \, , \, 2.3$ and defect width $\xi_\mathrm{s} = 0.07 \, , \, 0.06 \, , \, 0.05$ (corresponding to $\xi_\mathrm{s}/\xi_\mathrm{h} = 0.3 \, , \, 0.3 \, , \, 0.17$). The resultant far-from-equilibrium state is held for variables times up to $t \simeq 1~\mathrm{s}$, during which the universal dynamics develops and takes place.

\subsection{Measurement of density and momentum distributions}

The density and momentum distribution of the gas are measured after finite time of flight for $t_\mathrm{tof} = 1.5~\mathrm{ms}$ and $t_\mathrm{tof} = 46~\mathrm{ms}$ of free expansion. This gives access to the in-situ (iS) and time-of-flight (tof) density profiles, for which the atoms are detected through absorption and fluorescent imaging in a thin light sheet, respectively. We afterwards calculate the radially centered and integrated, density profiles in longitudinal direction. We correct the profiles for possible random sloshing effects. The quench and measurement is repeated for each experimental shot and hold time $t$ for $10 - 15$ times for the in-situ data and $25 - 50$ times for the time-of-flight data. The fast expansion in radial direction dilutes the gas and leads to ballistic expansion in the longitudinal direction. Since therefore the momentum of the particles during the expansion is approximately conserved, the density distribution after expansion converges to the in-situ momentum distribution of the cloud. We checked the effects of a finite dilution time via numerical simulations of the Gross-Pitaevskii equation using hydrodynamic models to determine the time dependence of the interaction constant $g$ for early times of the expansion. For the parameters of the experiment we could not find any significant deviations from a completely ballistic expansion in the longitudinal direction.

The pulled back momentum distribution converges for high $k$-values rapidly towards the true momentum distribution of the gas. For low momenta the finite in-situ size of the cloud does not allow for a clear separation of different momentum modes and atoms of different momenta overlap in the measured density after time-of-flight. This means that for a cloud of size $R$, particles with momentum $k\lesssim k_\mathrm{iS} = R m / (\hbar t_\mathrm{tof})$ do not have time to propagate sufficiently far outside the in-situ bulk density to be clearly separated. Therefore the pulled back momentum distribution for $k \lesssim k_\mathrm{iS}$ rather resembles the in-situ density profile than the actual momentum distribution of the gas.

\subsection{Scaling analysis}

We extract the universal scaling exponents $\alpha \, , \, \beta$ through a least-square fit of the analytical prediction \eq{ScalingFunction}, minimising 
\begin{align} \label{eq:chi2_rescaling}
 \chi^2(\alpha,\beta) = \frac{1}{N_t^2} \sum_{t,t_0}^{N_t} \chi^2_{\alpha,\beta}(t,t_0) ~,
\end{align}
where we average over all times $t$ and reference times $t_0$ within the scaling period. The local $\chi^2_{\alpha,\beta}(t,t_0)$ is calculated via
\begin{align}
 \chi^2_{\alpha,\beta}(t,t_0) &= \int_{k_\mathrm{l}}^{k_\mathrm{h}} \mathrm{d}k \left[ \frac{\left( (t/t_0)^{\alpha} \tilde{n}[(t/t_0)^{\beta} k,t_0] - \tilde{n}[k, t] \right)^2}{\tilde{\sigma}[(t/t_0)^{\beta} k,t_0]^2 + \tilde{\sigma}[k,t]^2} \right] \notag ~,
\end{align}
where $\sigma$ denotes the standard error of the mean and $\tilde{n}(k,t) = n(k,t)/N(t)$ and $\tilde{\sigma}(k,t) = \sigma(k,t)/N(t)$ are normalised by the total atom number to minimise the influence of atom loss during the evolution. Note however, that the atom loss is negligible during the time period where the system shows scaling behaviour. For later times the atom loss is $\approx 10\%$ per $100 \, \mathrm{ms}$, with a final atom number of approximately $40\%$ at the end of the evolution. The rescaling of the momentum variable inevitably leads to the necessity to compare the momentum distributions at momenta lying between the discrete values measured in the experiment. We therefore use a linear interpolation of the spectrum and its error at the reference time $t_0$ which allows to evaluate the experimental spectrum at all momenta $(t/t_0)^{\beta} k$. For the scaling analysis we symmetrise the spectrum by averaging the momentum distribution over $\pm k$ to lower the noise level.

Estimation of the exponents and their error is done via the likelihood function. In order to decouple the two exponents we take $\alpha = \beta + \Delta_{\alpha \beta}$ and fit the deviation $\Delta_{\alpha \beta}$ of the exponent from the theoretical expectation $\alpha \equiv \beta$. We therefore define the likelihood function
\begin{align}
 L(\Delta_{\alpha \beta},\beta) = \exp \left[ - \frac{1}{2} \chi^2(\Delta_{\alpha \beta}, \beta) \right] ~.
\end{align}
The most probable exponents are determined by the maximum of the likelihood function. The error of the estimate is determined by integrating the two-dimensional likelihood function along one dimension, and extracting the variance of the remaining exponent through a Gaussian fit to the marginal likelihood functions. We find excellent agreement between the marginal likelihood functions and the Gaussian fits. Therefore the Gaussian estimate of the error is equivalent to the (asymmetric) estimate using a change in the log-likelihood function by $1/2$. The reason for this good agreement is the above mentioned decoupling of the exponent, which results to a good degree in a two-dimensional, Gaussian likelihood function for $L(\Delta_{\alpha \beta},\beta)$. The estimates of the scaling exponents for different reference times are calculated equivalently (neglecting the sum over $t_0$ in \Eq{chi2_rescaling}). The estimate is insensitive to the upper cutoff $k_\mathrm{h}$ (within reasonable limits). The momentum $k_\mathrm{S}$, limiting the scaling region in the UV, is determined as the characteristic scale for which the mean deviation of the rescaled momentum distributions for $|k| \leq k_\mathrm{S}$ and averaged over all times $t$ in the scaling period exceeds the $95 \%$ confidence interval at the reference time $t_0$. The lower cutoff is taken as $k_\mathrm{l} = 0$. Excluding momenta $|k|<k_\mathrm{iS}$ leads to a small shift of the exponents towards lower values, but agrees well within the estimated errors of the exponents ($\lesssim 0.3 \sigma$ deviation). The results of the scaling analysis for three independent experimental realisations is shown in \FigsED{ED_rescaling}{ED_exponent_trev}. We find similar results in all cases. The exponents and errors reported in the main text, are estimated through the combined likelihood function $L = \prod_{i} L_i$, where $i$ labels the independent experimental realisations.

The universal function $f_\mathrm{S}$ is determined equivalently, where for each fixed exponent $\zeta$ the non-universal scales are determined through a least-square fit to each experimental realisation separately. To minimise the influence of the finite expansion of the gas, we consider momenta $|k| > k_\mathrm{iS}$ for the determination of $f_\mathrm{S}$. The likelihood function is subsequently defined by the averaged residuals of the scaled data as compared to the universal scaling function $f_\mathrm{s} = (1+k^\zeta)^{-1}$ for all realisations simultaneously . The error is again estimated using a Gaussian fit to the (one-dimensional) likelihood function. The non-universal scales for the most-likely exponent $\zeta = 2.39 \pm 0.18$ for the experimental realisations 1 to 3 are the characteristic momentum scale $k_0 = 2.61 \, , \, 2.28 \, , \, 3.97 \, \mu \mathrm{m}^{-1}$ and the norm $\mathcal{N} = 0.14 \, , \, 0.15 \, , \, 0.10$, respectively.

\subsection{Global observables}

We define the global observables
\begin{align} \label{eq:scaling_global_obs}
 \bar{N} &= \int\displaylimits_{|k| \leq (t/t_0)^{-\beta} k_0} \mathrm{d}k ~ \frac{n(k,t)}{N(t)}  \qquad \sim (t/t_0)^{- \Delta_{\alpha \beta}}\\
 \bar{M}_{n \geq 1} &= \int\displaylimits_{|k| \leq (t/t_0)^{-\beta} k_0} \mathrm{d}k ~ \frac{|k|^n n(k,t)}{{N\bar{N}(t)}}  ~\sim (t/t_0)^{- n \beta} ~,
\end{align}
where $k_0 = 6.5 \dots 8 \imum$ defines the high-momentum cutoff for the scaling region in $k$. We consider, in the main text, the fraction of particles in the scaling region $\bar{N} \sim (t/t_0)^{\Delta_{\alpha \beta}}$ and the mean kinetic energy per particle in the scaling region $\bar{M}_2 \sim (t/t_0)^{-2 \beta}$. Note that the global observables $\bar{N}$ and $\bar{M}$ show independent scaling in time with the exponents $\Delta_{\alpha \beta}$ and $\beta$, while we highlight that the integral ranges depend non-trivially on $\beta$. The results for each experimental realisation are shown in \FigED{ED_global_observables}. In the main text we report the result obtained by averaging over all experimental realisations.


\subsection{Model fits}
The density profile $\rho(z)$ is determined by a fit to the experimental in-situ density, measured after $\tau = 1.5~\mathrm{ms}$ of free expansion. In case of the RDM we consider for a fixed atom number $N(t)$ a scaled density profile $\rho[z,t] = b^{-1}(t) \rho[z/(b(t)]$ in the Thomas-Fermi approximation, leaving the scaling factor $b(t)$ as the only free parameter (more precisely the result of Gerbier \cite{Gerbier2004} is used for the density profile, which takes the radial swelling of the condensate into account). We neglect possible finite temperature fluctuations as well as contribution of radially excited states in the RDM, assuming the gas to be dominated by solitonic defects. For early times, the high momentum modes do not show a significant thermal occupation, and we find good accordance with the RDM. 

In case of the QC, we determine the thermal density profile for a given temperature $T$ and chemical potential $\mu$, through simulations of the stochastic Gross-Pitaevskii equation (see e.g.~\cite{Blakie2008b}). The broadening of the density distribution is herein due to the finite temperature of the gas. The density profile is subsequently fitted via $\rho(z,t) = \rho_\mathrm{QC}[z,T(t),\mu(t)] + \rho_\perp[z,T(t),\mu(t)]$. Here we take into account the thermal occupation of radially excited states $\rho_\perp$ within the semiclassical approximation, which are non-negligible for late times. The chemical potential $\mu$ is fixed by the total atom number, through $\int \mathrm{d}z \, \rho(z,t) = N(t)$.

The fitted density profiles are subsequently used to determine the single-particle momentum distribution $n(k,t)$ of the inhomogeneous system by a least-square fit of the experimental data to the theoretical predictions within the local density approximation. For both models we restrict the fitting region to $|k|>k_\mathrm{iS}$, due to the simplified hydrodynamic model for the finite expansion of the gas. The RDM \cite{Schmidt:2012kw} is fitted over the full momentum range accessible in the experiment. For high defect densities the RDM fit shows correlations between the determined defect density and width, since these two scales become of the same order for the observed far-from equilibrium state. As it is theoretically expected that the defect width is approximately conserved during the evolution, we fix the defect width to its mean value within the first $25 \, \mathrm{ms}$ of the evolution, leaving the defect density as the only free parameter. We find reasonable agreement between the RDM results and the independent scaling analysis. In particular, the RDM is clearly preferred as compared to a thermal distribution within the scaling period.

For the fits in thermal equilibrium we consider a quasicondensate model \cite{richard2003momentum}, including thermal occupation of radially excited states \cite{davis2012yang}. Considering the validity regime of the QC model, we restrict the fitting procedure to momentum modes with energy less then $\hbar \omega_\perp$. We consequently determine the chemical potential $\mu$, by fixing the atom number within this region of momentum space. This leads to a slight shift in the chemical potential as compared to the in-situ fits. For late times we find excellent agreement to the experimental data, showing the relaxation of the system to thermal equilibrium.


\onecolumngrid


\vspace*{1cm}

\begin{figure*}[h]
\begin{center}
\includegraphics[width=0.6\textwidth]{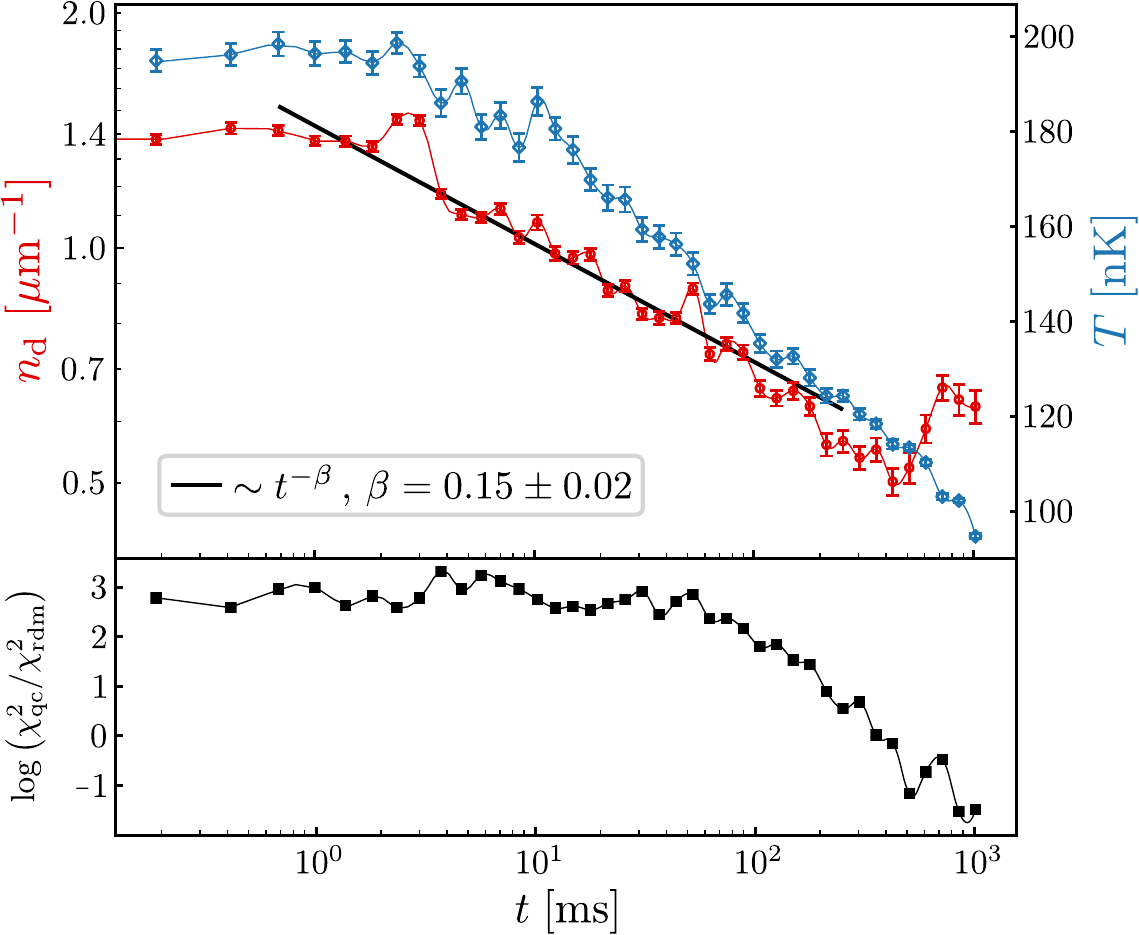}
\end{center}
\caption{{\bf Results of random-defect and quasicondensate models.}
Time evolution of the characteristic scales for the experimental data scan 1 reported in the main text. The resulting temperature (blue, log-lin scale) and defect density (red, log-log scale) are shown in the upper panel for the full time evolution. The defect width for the RD model is fixed to $\xi_d = 0.087 \, \mu\mathrm{m}$, determined by the mean over the first $25 ~\mathrm{ms}$ during the evolution. The defect density within the scaling region shows power-law dependence consistent with the determined scaling exponent of the self-similar evolution. For later times deviations occur, signaling the end of the scaling region. The quality of the model fit is depicted in the lower panel (black squares), where a positive (negative) value favors the RD (QC) model. The RD model is strongly preferred for $t \lesssim 100 \mathrm{ms}$, beyond which the system converges to a thermal QC within $\approx 400 ~\mathrm{ms}$. The absolute values of the reduced $\chi^2$ for the RD (QC) model are $\approx 1$ ($\approx 25$) and $\approx 5$ ($\approx 1$) for early and late times, respectively.
}
\label{fig:ED_fitresults}
\end{figure*}

\begin{figure*}[thb!]
\begin{center}
\includegraphics[width=0.75\textwidth]{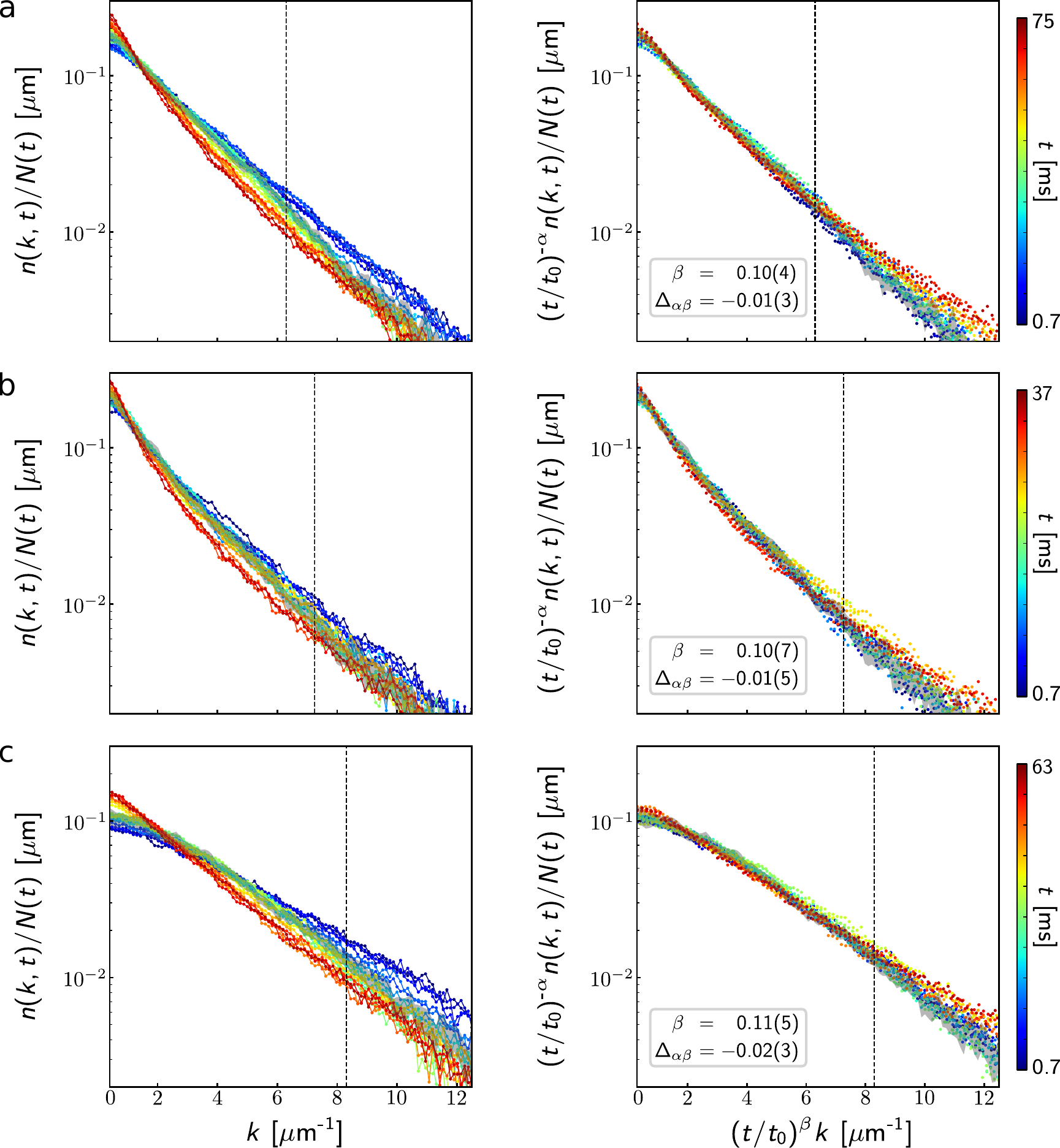}
\end{center}
\caption{{\bf Rescaling analysis for different initial conditions.}
{\bf a} to {\bf c,} original (left row) and rescaled (right row) single-particle momentum distribution for scans 1 to 3 reported in the main text. The time is encoded in colors from blue to red. The gray dashed vertical line indicates the scaling regime in $k$. The scaling exponents are in excellent agreement with the reported mean values in the main text. Note that here we compare the data for the full experimental resolution in $k$. The distribution at the reference time is given by the gray line, its width marking the $95\%$ confidence interval.
}
\label{fig:ED_rescaling}
\end{figure*}

\begin{figure}[thb!]
\centering
\includegraphics[width=1\textwidth]{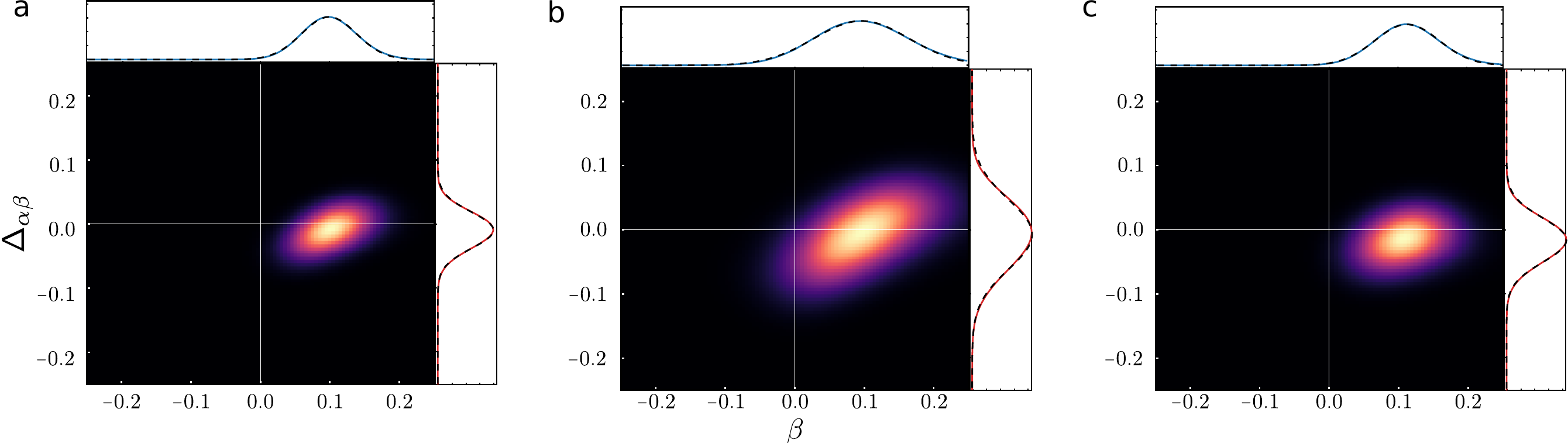}
\caption{{\bf Likelihood function for different initial conditions.}
{\bf a} to {\bf c,} two-dimensional likelihood functions and marginal likelihood functions for scans 1 to 3 reported in the main text. A clear peak at non-zero $\alpha \approx \beta$ is visible for each realisation. Note that for scan 2 a small condensate may have been present before the quench, which leads to the larger extent of the likelihood function. Gaussian fits to the marginal likelihood functions are in excellent agreement and determine the error of the scaling exponents reported in \FigED{ED_rescaling}.
}
\label{fig:ED_likelihood}
\end{figure}

\begin{figure*}[thb!]
\centering
\includegraphics[width=0.65\textwidth]{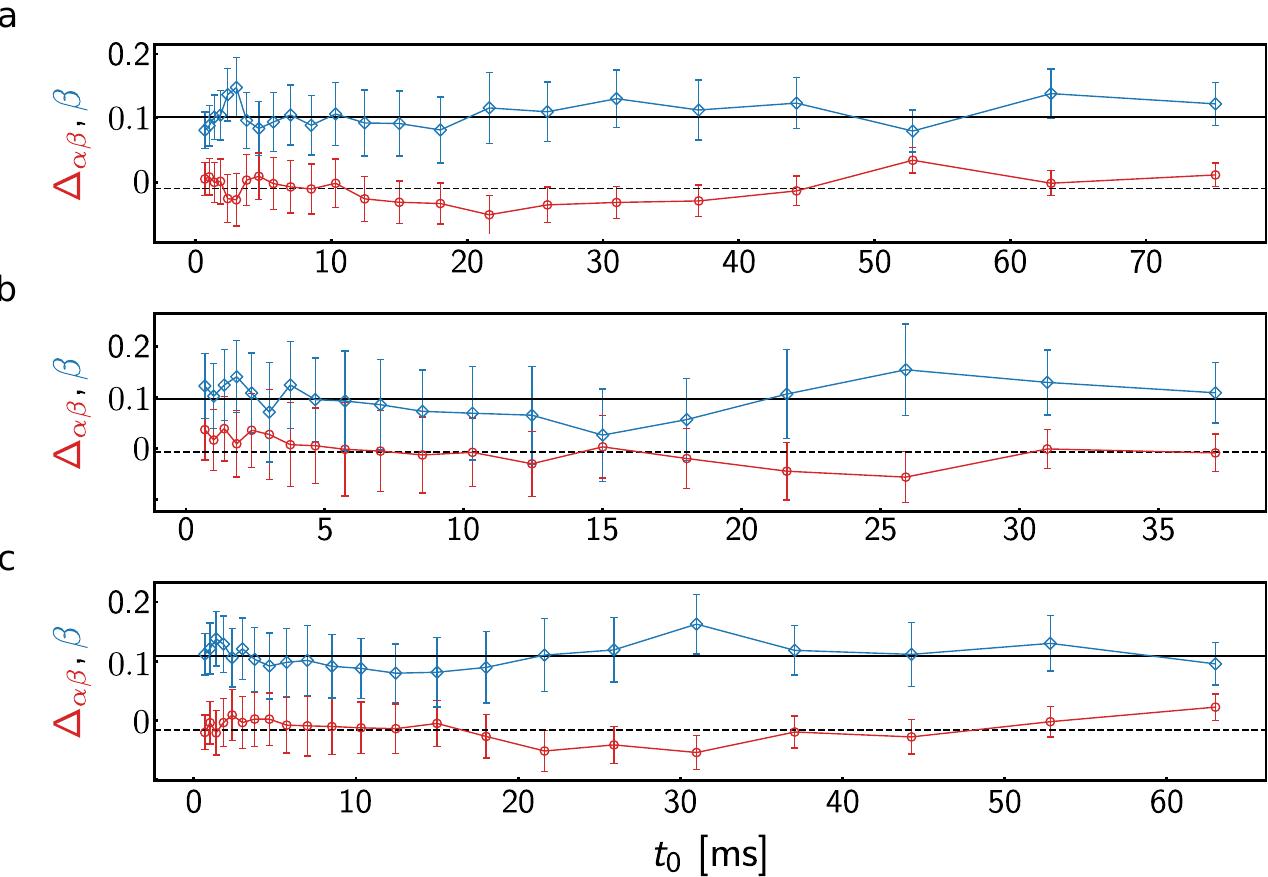}
\caption{{\bf Time evolution of scaling exponents for different initial conditions.}
{\bf a} to {\bf c,} scaling exponents for scans 1 to 3 determined from the likelihood function for each reference time $t_0$ are in good agreement with the predicted mean. The errors denote the standard deviation obtained by a Gaussian fit to the marginal likelihood functions at each reference time separately.
}
\label{fig:ED_exponent_trev}
\end{figure*}

\begin{figure*}[thb!]
\centering
\includegraphics[width=1\textwidth]{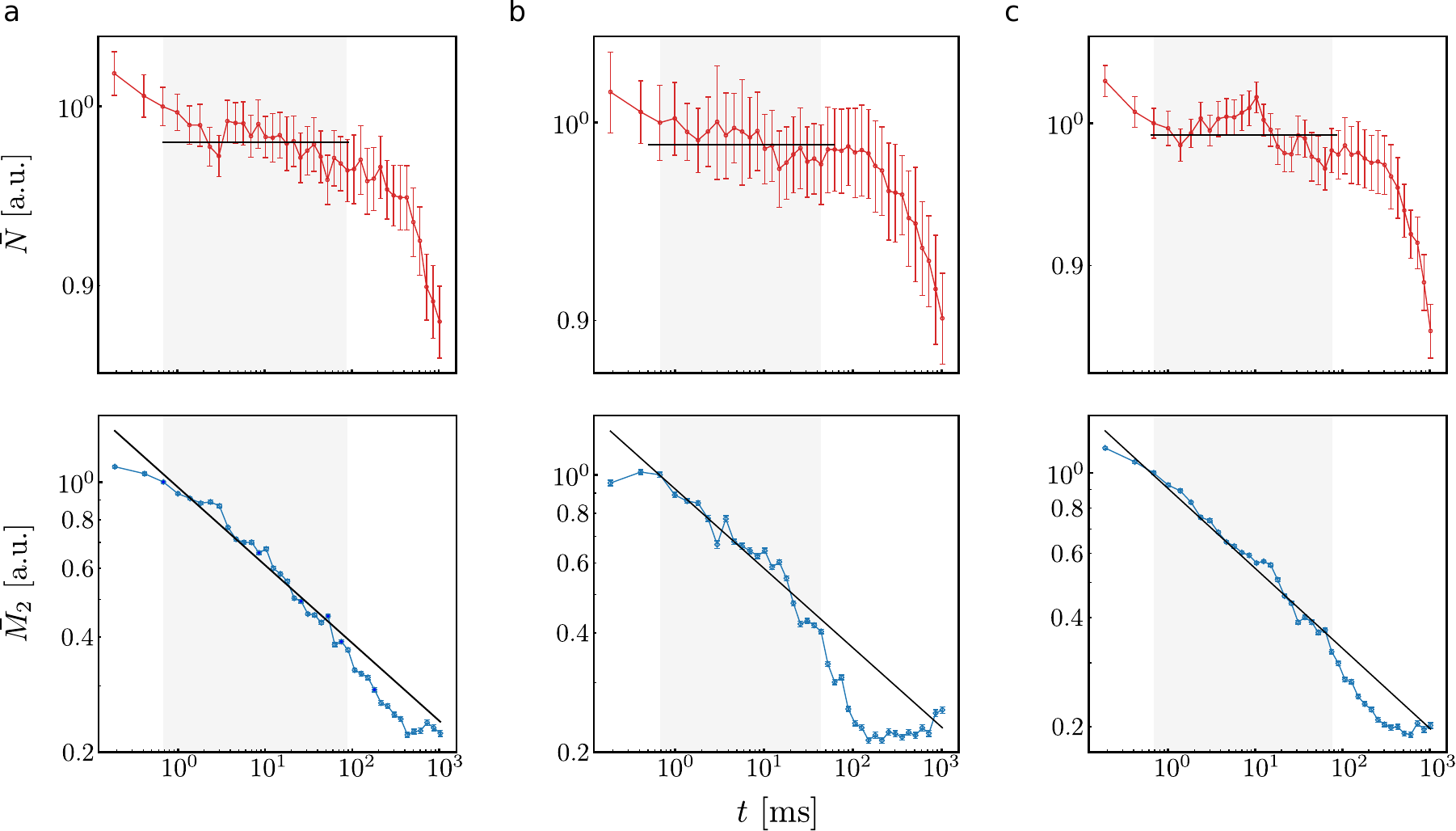}
\caption{{\bf Spatially averaged observables for different initial conditions.}
{\bf a} to {\bf c,} time evolution of averaged observables $\bar{N}$ and $\bar{M}_2$ for the scans 1 to 3. Within the scaling region (shaded gray area) $\bar{N}$ is approximately conserved. The solid black lines are the approximately conserved value and scaling solutions \eq{scaling_global_obs}, respectively. The errors mark the $95 \%$ confidence interval.
}
\label{fig:ED_global_observables}
\end{figure*}

\end{document}